\documentclass[aps,prl,twocolumn,showpacs,byrevtex]{revtex4}

\usepackage{times,mathptm}
\usepackage[dvips]{graphicx,color}

\begin{document}

\title{Origin of the Metal-Insulator Transition of Indium Atom Wires on Si(111)}
\author{Sun-Woo Kim and Jun-Hyung Cho$^{*}$}
\affiliation{Department of Physics and Research Institute for Natural Sciences, Hanyang University,
17 Haengdang-Dong, Seongdong-Ku, Seoul 133-791, Korea}

\date{\today}

\begin{abstract}
As a prototypical one-dimensional electron system, self-assembled indium (In) nanowires on the Si(111) surface have been believed to drive a metal-insulator transition by a charge-density-wave (CDW) formation due to electron-phonon coupling. Here, our first-principles calculations demonstrate that the structural phase transition from the high-temperature 4${\times}$1 phase to the low-temperature 8${\times}$2 phase occurs through an exothermic reaction with the consecutive bond-breaking and bond-making processes, giving rise to an energy barrier between the two phases as well as a gap opening. This atomistic picture for the phase transition not only identifies its first-order nature but also solves a long-standing puzzle of the origin of the metal-insulator transition in terms of the ${\times}$2 periodic lattice reconstruction of In hexagons via bond breakage and new bond formation, not by the Peierls instability-driven CDW formation.
\end{abstract}


\maketitle


Low-dimensional electronic systems are of great interest in contemporary condensed-matter physics because of their susceptibility to charge density wave (CDW) instability~\cite{Carpinelli}, non-Fermi liquid behavior~\cite{Blumenstein}, spin ordering~\cite{Erwin,Li}, and superconductivity~\cite{Qin,TZhang} at low temperatures. Specifically, metal-atom adsorption on semiconductor surfaces provides a unique playground for the exploration of such exotic physical phenomena~\cite{Tejeda,Snijders}. We here focus on a prototypical example of quasi-one-dimensional (1D) systems, self-assembled indium (In) atom wires on the Si(111) surface~\cite{Bunk,Yeom1,Cho}. Each In wire consists of two zigzag chains of In atoms [see the left panel of Fig. 1(a)]~\cite{Bunk}. Below ${\sim}$120 K, this quasi-1D system undergoes a reversible phase transition initially from a 4${\times}$1 structure to a 4${\times}$2 one, then to an 8${\times}$2 structure~\cite{Yeom1,Kumpf}, showing a period doubling both parallel and perpendicular to the In wires. This (4${\times}$1)${\rightarrow}$(8${\times}$2) structural phase transition is accompanied by a metal-insulator (MI) transition~\cite{Yeom1,Ahn,Yeom2}. For the explanation of such a MI transition, the CDW mechanism due to a Peierls instability was initially proposed~\cite{Yeom1,Ahn,Yeom2,Park}, but subsequently other mechanisms based on an order-disorder transition~\cite{Gonzalez1,Gonzalez2} and many-body interactions~\cite{GLee} have been proposed. Despite such debates, the CDW mechanism has been most widely believed to drive the observed MI transition~\cite{Yeom1,Ahn,Yeom2,Park,Morikawa,SHUhm,DMOh}. It is noted that the CDW formation invokes the strong coupling between lattice vibrations and electrons near the Fermi level $E_F$, caused by Fermi surface nesting with a nesting vector 2$k_{F}$ = ${\pi}$/$a_{\rm x}$ ($a_{\rm x}$: the 4${\times}$1 lattice constant along the In wires)~\cite{Yeom1,Gruner}. The resulting Peierls dimerization was believed to occur on each chain, and the two dimerized chains further interact with each other, leading to a coupled double Peierls-dimerized chain model~\cite{Cheon} [see Fig. 1(b)].

\begin{figure}[ht]
\centering{ \includegraphics[width=7.0cm]{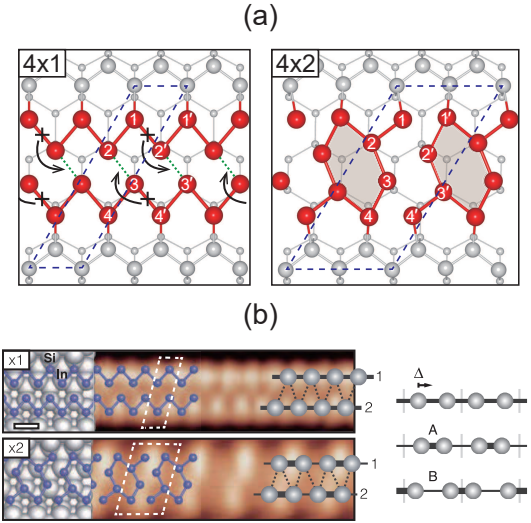} }
\caption{(Color online) (a) Top view of the optimized 4${\times}$1 (left panel) and 4${\times}$2 (right panel) structures. The dark and gray circles represent In and Si atoms, respectively. For distinction, Si atoms below In chains are drawn with small circles. Each unit cell is indicated by the dashed line. The STM images of the 4${\times}$1 and 4${\times}$2 structures is displayed in (b), together with the overlap of the coupled double Peierls-dimerized chain model (from Ref.~\cite{Cheon}): Schematic of two dimerized phases, A (with a positive displacement ${\Delta}$) and B (with a negative ${\Delta}$), of a single Peierls chain is also given on the right side. }
\end{figure}

Regarding the phase transition of the In/Si(111) system, there are still some unsettled issues. Although it is well established that the structural model of the 8${\times}$2 phase is constituted by the basic building block of In ``hexagon" [see the right panel of Fig. 1(a)], the microscopic mechanism of the hexagon formation is so far unclear whether it is driven by the Peierls dimerization on the two chains~\cite{Yeom1,Cheon} or by a shear distortion where neighboring chains are displaced in opposite directions~\cite{Gonzalez1}. Recently, it was also reported that both the covalent bonding and van der Waals (vdW) interactions between the two chains play crucial roles in forming In hexagons~\cite{HJKim}. Moreover, the order of the phase transition has been controversial whether it belongs to first-order~\cite{Park2,Wall,Klasing} or second-order~\cite{Gonzalez1,Gonzalez2,Guo,GLee2}. According to the mean-field theory~\cite{Gruner}, the CDW or order-disorder mechanism can be classified as the second-order phase transition. However, existing scanning tunneling microscopy (STM) experiments have reached the conflicting conclusions between the first-order~\cite{Park2} and second-order~\cite{Guo,GLee2} transitions, whereas a recent high-resolution low-energy electron diffraction (HRLEED) study~\cite{Klasing} observed a robust hysteresis of diffraction spot intensities as the sample temperature slowly increases and decreases during the (4${\times}$1)${\leftrightarrow}$(8${\times}$2) phase transition. The latter observation~\cite{Klasing} obviously indicates the existence of energy barrier between the two phases, representing the first-order transition~\cite{Landau}. All of these controversies on the microscopic mechanism of the hexagon formation and the order of the phase transition reflect our incomplete understanding of the origin of the phase transition in the In/Si(111) system.

In this paper, using first-principles density-functional theory (DFT) calculations, we propose the atomistic picture for the phase transition of the In/Si(111) system to resolve the existing problems such as the microscopic mechanism of the hexagon formation, the order of the phase transition, as well as the origin of the MI transition. We find that the hexagon formation is driven by an exothermic reaction with the consecutive bond-breaking and bond-making processes, giving rise to an energy barrier as well as a gap opening. This atomistic picture not only reveals the first-order nature of the (4${\times}$1)${\leftrightarrow}$(8${\times}$2) phase transition, but also illustrates how the observed solitons~\cite{Zhang,THKim,Cheon} can be created at the boundary of two differently oriented hexagon structures. Our findings clarify that the MI transition of the In/Si(111) system is attributed to the ${\times}$2 periodic lattice reconstruction of In hexagons via the bond breakage and the new bond formation, rather than the CDW formation arising from Fermi surface nesting~\cite{Yeom1,Ahn,Yeom2,Park,Cheon}.

To properly predict the energetics of the 4${\times}$1, 4${\times}$2, and 8${\times}$2 structures~\cite{HJKim,Zhang2,SWKim}, we have performed the van der Waals energy corrected~\cite{vdW} hybrid DFT calculations using the FHI-aims~\cite{Aims} code that gives an accurate, all-electron description based on numeric atom-centered orbitals, with ``tight" computational settings. For the exchange-correlation energy, we employed the hybrid functional of HSE06~\cite{HSE1,HSE2}. The ${\bf k}$-space integrations in various unit-cell calculations were done equivalently with 64 ${\bf k}$ points in the surface Brillouin zone of the 4${\times}$1 unit cell. The Si(111) substrate (with the Si lattice constant $a_0$ = 5.418 {\AA}) below the In chains was modeled by a 6-layer slab with ${\sim}$30 {\AA} of vacuum in between the slabs. Each Si atom in the bottom layer was passivated by one H atom. All atoms except the bottom layer were allowed to relax along the calculated forces until all the residual force components were less than 0.01 eV/{\AA}.

\begin{figure}[ht]
\centering{ \includegraphics[width=7.0cm]{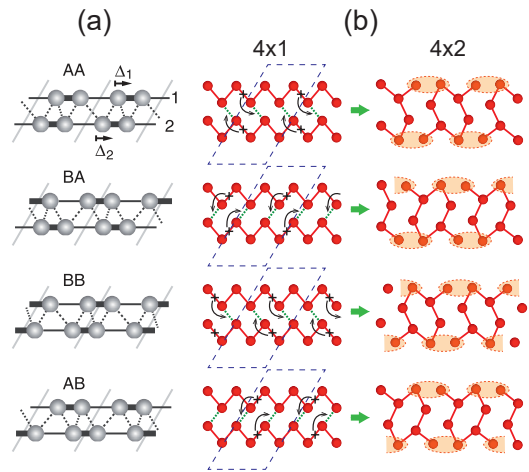} }
\caption{(Color online) (a) Schematics of four degenerate CDW ground phases (denoted as AA, BA, BB, and AB) in a coupled double Peierls-dimerized chain model (from Ref.~\cite{Cheon}). (b) Atomistic picture for the formation of the four degenerate hexagon structures. The dashed lines in (b) indicate the 4${\times}$2 unit cell. For comparison with the Peierls dimerization in (a), the filled ellipses (or half-ellipses) are drawn in the outer In atoms of the 4${\times}$2 structure in (b). }
\end{figure}

\begin{figure*}[ht]
\centering{ \includegraphics[width=17.0cm]{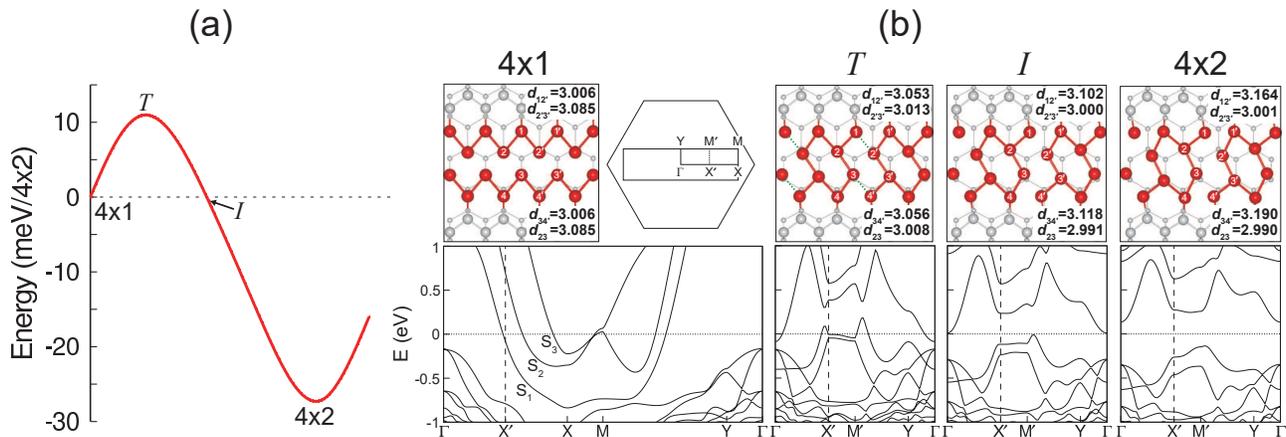} }
\caption{(Color online) (a) Calculated energy profile along the transition path from the 4${\times}$1 to the 4${\times}$2 structure. (b) Atomic geometries and surface band structures of the 4${\times}$1 structure, the $T$ state, the $I$ state, and the 4${\times}$2 structure. The surface Brillouin zones of the 4${\times}$1 and 4${\times}$2 structures are also drawn. The In-In interatomic distances (in {\AA}) are given. The energy zero in (b) represents the Fermi level.}
\end{figure*}

According to the previously proposed CDW mechanism~\cite{Yeom1,Ahn,Yeom2,Park,Cheon} of the structural phase transition, the ground state of the In wire was described by a double Peierls-dimerized chain model with a zigzag interchain coupling between two chains [see Fig. 1(b)]: that is, below ${\sim}$120 K, the In wire undergoes a period doubling CDW transition through the Peierls dimerization on both chains. Since there are the two dimerization directions that form ``A" phase with a positive displacement ${\Delta}$ and ``B" phase with a negative ${\Delta}$ [see Fig. 1(b)] on the two chains [i.e., chain ``1" and chain ``2" in Fig. 1(b)], the double chain can have four degenerate CDW ground phases, which were denoted as AA, BA, BB, and AB in Ref.~\cite{Cheon} [see Fig. 2(a)]. It is, however, noticeable that such a coupled double Peierls-dimerized atomic chain is unlikely to be fit well for the well-established hexagon structural model~\cite{Gonzalez1} which involves the bond formation between the two chains. Therefore, we here propose a new atomistic model of the hexagon formation, where the bonds on the chains 1 and 2 are broken to form the new bonds between both chains. For one example of such bond breakage and new bond formation, the bond $b_{12'}$ between In$_1$ and In$_{2'}$ atoms on chain 1 and the bond $b_{34'}$ between In$_3$ and In$_{4'}$ atoms on chain 2 are broken to form the new bonds $b_{2'3'}$ and $b_{23}$, respectively, leading to the formation of the 4${\times}$2 hexagon structure [see Fig. 1(a)]. Note that the bond breaking/forming of $b_{12'}$/$b_{2'3'}$ accompanies that of $b_{34'}$/$b_{23}$ and vice versa, as discussed below. Since there are four different bonds on chain 1 (or 2) within the 4${\times}$2 unit cell, four degenerate hexagon structures can be generated as shown in Fig. 2(b), which correspond to AA, BA, BB, and AB configurations in Ref.~\cite{Cheon}.

The present atomistic model of the hexagon formation is expected to have an energy barrier for the consecutive bond-breaking and bond-making processes. To find this energy barrier, we calculate the energy profile along the transition path between the 4${\times}$1 and 4${\times}$2 structures by using the nudged elastic-band method~\cite{NEB}. The calculated energy profile is displayed in Fig. 3(a). We find that the transition ($T$) state is higher in energy than the 4${\times}$1 structure by 5.4 meV per 4${\times}$1 unit cell, yielding an energy barrier ($E_b$) of ${\sim}$11 meV on going from the 4${\times}$1 to the 4${\times}$2 structure. Since the 4${\times}$2 structure is more stable than the 4${\times}$1 structure by 13.6 meV per 4${\times}$1 unit cell (see Table I), we can say that the hexagon formation occurs through an exothermic reaction with bond breakage and new bond formation. It is noted that the 8${\times}$2 structure is further stabilized over the 4${\times}$1 structure by 32.9 meV per 4${\times}$1 unit cell (see Table I), and $E_b$ is reduced to be ${\sim}$8 meV on going from 4${\times}$1 to 8${\times}$2, which is much smaller than that ($E_b$ = 40 meV) obtained by a previous DFT calculation with the local-density approximation~\cite{Wall}. The existence of the energy barrier for the (4${\times}$1)${\rightarrow}$(8${\times}$2) phase transition is consistent with a recent HRLEED experiment~\cite{Klasing} where the energy barrier was confirmed by the observation of a hysteresis of diffraction spot intensities upon slow increase and decrease of the sample temperature at the (4${\times}$1)${\leftrightarrow}$(8${\times}$2) phase transition. Thus, the present theory and the previous HRLEED experiment support the first-order nature of the phase transition, contrasting with the second-order nature deduced from the CDW~\cite{Guo,GLee2} or order-disorder mechanism~\cite{Gonzalez1,Gonzalez2}.

Figure 3(b) shows the atomic geometries of the $T$ state and an intermediate ($I$) state, together with those of the 4${\times}$1 and 4${\times}$2 structures. We find that, along the transition path from the 4${\times}$1 to the 4${\times}$2 phase, the In-In interatic distance $d_{12'}$ ($d_{34'}$) increases as 3.01 (3.01), 3.05 (3.06), 3.10 (3.12), and 3.16 (3.19) {\AA} for 4${\times}$1, $T$, $I$, and 4${\times}$2, respectively, while $d_{2'3'}$ ($d_{23}$) decreases as 3.09 (3.09), 3.01 (3.01), 3.00 (2.99), and 3.00 (2.99) {\AA} [see Fig. 3(b)]. These results indicate that during the structural phase transition the bond breakage of $b_{12'}$ proceeds the new bond formation of $b_{2'3'}$, simultaneously taking place with the bond breakage of $b_{34'}$ and the new bond formation of $b_{23}$. It is noticeable that such bond-breaking and bond-making processes leading to the hexagon formation were not taken into account in the coupled double Peierls-dimerized chain model~\cite{Cheon}. Thus, unlike the present atomistic model of the hexagon formation, the latter model~\cite{Cheon} does not properly describe the hexagon structure of the indium wire.

\begin{table}[ht]
\caption{Calculated total energies of the 4${\times}$2 and 8${\times}$2 structures relative to the 4${\times}$1 structure, together with the band gap $E_{\rm g}$. }
\begin{ruledtabular}
\begin{tabular}{lcc}
& 4${\times}$2  & 8${\times}$2   \\ \hline
${\Delta}E$ (meV per 4${\times}$1 unit cell)                     &       --13.6     &     --32.9      \\
$E_{\rm g}$ (eV)                    &       0.27     &  0.31  \\
\end{tabular}
\end{ruledtabular}
\end{table}

The calculated surface band structures along the transition path between the 4${\times}$1 and 4${\times}$2 structures are also displayed in Fig. 3(b). We find that the 4${\times}$1 structure has the three surface bands $S_1$, $S_2$, and $S_3$ crossing the Fermi level, showing a metallic feature. On the other hand, it is clearly seen that the band gap opens during the bond-breaking and bond-making processes. Specifically, the gap opening starts from the $I$ state, leading to a band gap ($E_{\rm g}$) of 0.27 (0.31) eV at the 4${\times}$2 (8${\times}$2) structure: see Fig. 3(b) (Fig. 5S of the Supplemental material~\cite{supp}). These values of $E_{\rm g}$ agree well with those (${\sim}$0.3 eV) obtained from surface transport measurements~\cite{Tanikawa} and scanning tunneling spectroscopy~\cite{Gonzalez2}. Based on a coupled double Peierls-dimerized chain model [see Fig. 1(b)], a tight-binding Hamiltonian analysis showed that the Peierls dimerization on both chains hybridizes the $S_1$ and $S_2$ states to produce a gap opening~\cite{Cheon}. However, this Peierls instability-driven gap opening is characteristically different from the present gap opening driven by the ${\times}$2 periodic lattice reconstruction of In hexagons that involves the bond-breaking and bond-making processes within the In wire. Since the former Peierls chain model~\cite{Cheon} is lacking in the atomic description of the hexagon structure as discussed above, we believe that the CDW mechanism would not be the origin of the MI transition of the In/Si(111) system.

\begin{figure}[ht]
\centering{ \includegraphics[width=7.0cm]{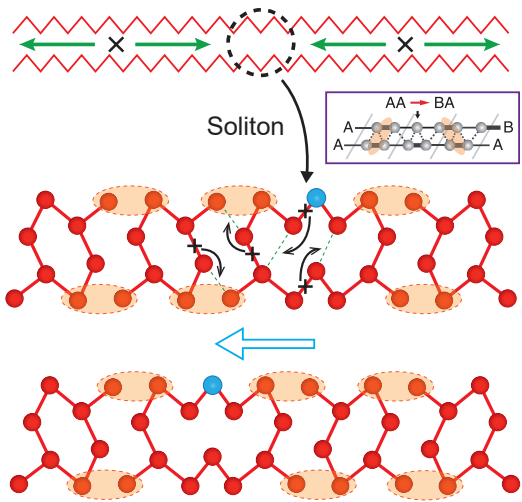} }
\caption{(Color online) Schematic diagrams of the soliton formation. In the top panel, the initial bond-breaking and bond-making positions are marked (X), and the arrows represent the propagations of In hexagons. In the inset, the soliton configuration AA${\rightarrow}$BA between the AA and BA CDW phases is from Ref.~\cite{Cheon}. The corresponding soliton structure is sketched based on the hexagon model, together with the possible bond-breaking and bond-making processes for the soliton movement. The kink atom indicated by the small arrow in the inset is drawn with bright color. }
\end{figure}

Finally, it is noteworthy that the coupled double Peierls chain model classified three types of topological edge states at the domain boundary between two different CDW phases~\cite{Cheon}: i.e., right-chiral, left-chiral, and nonchiral solitons among the total twelve possible solitons arising from four degenerate CDW phases. However, there remain the question of how the multiple CDW phases and the resulting solitons can be created from the 4${\times}$1 phase along the In wires. Based on the present atomistic picture for the phase transition, we speculate that around the phase transition temperature the bond breakage takes place at random sites of the 4${\times}$1 phase, initiating the hexagon formation which then propagates along the In wire (see sketch in the upper panel of Fig. 4). This speculation is supported by our calculation with a larger 4${\times}$8 supercell, where after the formation of a single hexagon it is converged to the 4${\times}$2 hexagon structure without any barrier. Indeed, the propagation of the hexagon structure was observed by an ultrafast time-resolved reflection high energy electron diffraction experiment~\cite{Wall} where adsorbates trigger the propagation of the phase front of the 8${\times}$2 structure with a constant velocity of 82 m/s. Therefore, a soliton is expected to be created at the midpoint between two bond-breakage positions after the propagation of two different hexagon structures [see Fig. 2(b)] in the opposite directions. Moreover, the created solitons can move along the In wire by the activation of bond-breaking and bond-making processes. For instance, one soliton structure corresponding to the AA${\rightarrow}$BA configuration in Ref.~\cite{Cheon} is sketched in Fig. 4, together with its movement. Future theoretical and experimental works are anticipated to investigate the detailed kinetics of solitons such as the activation energy for the soliton motion, the propagation speed of solitons, the transformation of soliton configurations, and so on.

In summary, based on first-principles DFT calculations, we have proposed the atomistic picture for the phase transition of the In/Si(111) system, where the low-temperature hexagon structure is formed through an exothermic reaction with the consecutive bond-breaking and bond-making processes. During such bond breakage and new bond formation, we found the existence of an energy barrier as well as a gap opening. Therefore, we revealed not only the first-order nature of the (4${\times}$1)${\leftrightarrow}$(8${\times}$2) phase transition but also the origin of the MI transition in terms of the ${\times}$2 periodic lattice reconstruction of In hexagon, rather than the prevailing Peierls instability-driven CDW mechanism. We anticipate that our findings will stimulate future theoretical and experimental works to reinterpret many of the existing controversial issues in the prototypical 1D In/Si(111) system in the light of the presently proposed atomistic picture for the phase transition.

This work was supported by National Research Foundation of Korea (NRF) grant funded by the Korea Government (MSIP) (2015R1A2A2A01003248). The calculations were performed by KISTI supercomputing center through the strategic support program (KSC-2015-C3-044) for the supercomputing application research. S.W.K. acknowledges support from POSCO TJ Park Foundation.

\noindent $^{*}$ Corresponding author: chojh@hanyang.ac.kr

\widetext
\clearpage

\makeatletter
\renewcommand{\fnum@figure}{\figurename ~\thefigure{S}}
\renewcommand{\fnum@table}{\tablename ~\thetable{S}}
\makeatother

\vspace{2.4cm}
{\bf \huge Supplemental Material}
\vspace{0.2cm}

\vspace{1cm}
{\Large 1. The atomic geometry and the surface band structure of the 8$\times$2 structure.} \\ \\

The atomic geometry and the surface band structure of the 8$\times$2 structure obtained using van der Waals energy corrected hybrid DFT calculations (HSE+vdW) are shown in Fig. 5S.

\begin{figure}[h]
\centering{ \includegraphics[width=16.0cm]{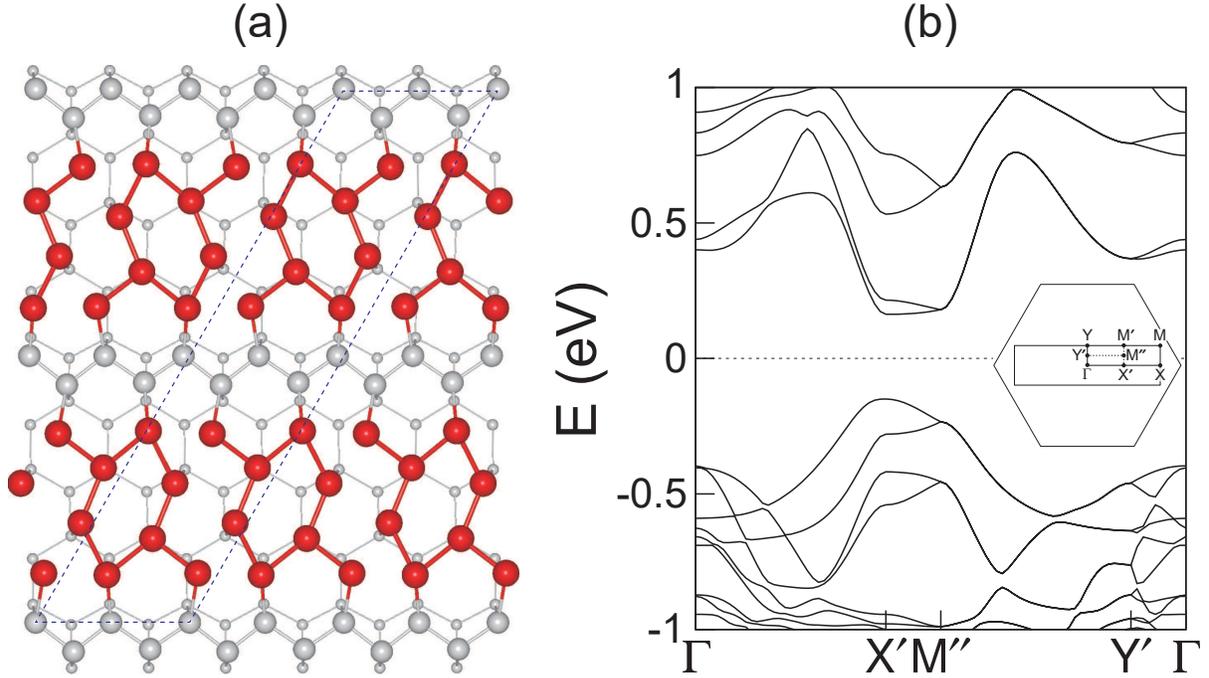}}
\caption{(a) Top view of the optimized 8$\times$2 structure. Unit cell is indicated by the dotted line. (b) The surface band structure of the 8$\times$2 structure obtained using HSE+vdW. The inset in (b) shows the surface Brillouin zone for the 4$\times$1, 4$\times$2, and 8$\times$2 unit cells within that for the 1$\times$1 unit cell. The energy zero represents the Fermi level.}
\end{figure}

\vspace{0.2cm}

\end{document}